# A Cautionary Note on the Thermal Boundary Layer Similarity Scaling for the Turbulent Boundary Layer

David Weyburne[1]
Air Force Research Laboratory
2241 Avionics Circle
WPAFB OH 45433

Wang and Castillo have developed empirical parameters for scaling the temperature profile of the turbulent boundary layer flowing over a heated wall in the paper X. Wang and L. Castillo, J. Turbul., **4**, 1(2003). They presented experimental data plots that showed similarity type behavior when scaled with their new scaling parameters. However, what was actually plotted, and what actually showed similarity type behavior, was not the temperature profile but the defect profile formed by subtracting the temperature in the boundary layer from the temperature in the bulk flow. We show that if the same data and same scaling is replotted as just the scaled temperature profile, similarity is no longer prevalent. This failure to show both defect profile similarity and temperature profile similarity is indicative of false similarity. The nature of this false similarity problem is discussed in detail.

## 1. Introduction

Prandtl introduced the concept a boundary layer for fluid flow past a solid over a hundred years ago [1]. The boundary layer concept for flow over a flat plate is depicted in Figure 1. As the flowing fluid impinges on the flat plate, a boundary layer is formed along the plate such that the velocity and temperature at the wall gradually transitions to the bulk values.

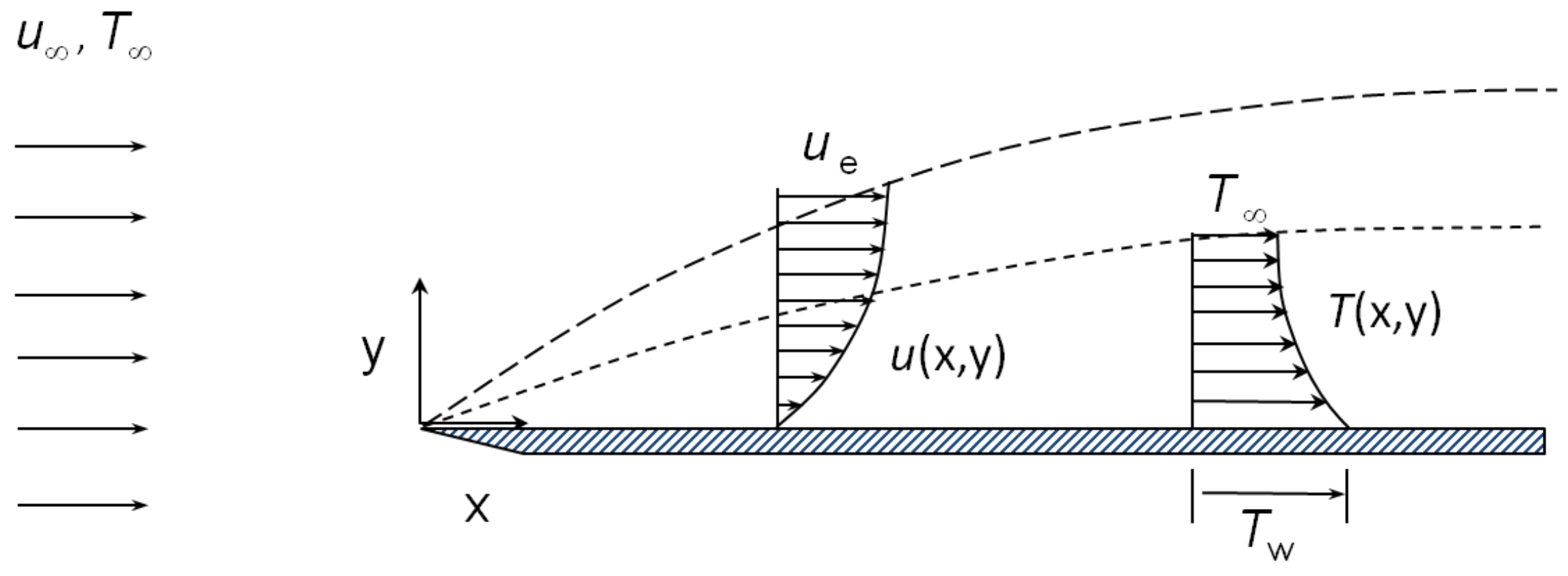

**Fig. 1: A schematic diagram showing the flat plate 2-D flow geometry and variables.**

For certain flow situations, the downstream velocity or temperature profile can be a simple stretched version of the upstream profile. For this situation the fluid is said to display similarity of the velocity (or temperature) profiles. It is one of the basic concepts of fluid flow theory going back to the pioneering dimensional analysis work of Reynolds [2] in the late 1800's. For similarity of the profiles depicted above, two temperature profile curves from different stations along the wall in the flow direction are said to be similar if they differ only by a scaling parameter in $y$ and a scaling parameter of the temperature profile $T(x,y)$. The temperature profile in this case is defined as the temperature taken at multiple *y* values starting from the wall moving outwards at a fixed *x* value.

[1]Email: David.Weyburne@gmail.com

For certain laminar flow cases, it is possible to determine the exact mathematical nature of these stretching parameters which provides valuable insights into the physical processes controlling the fluid and heat flow along the plate. For turbulent boundary flows this same path is not possible so what is usually done is to make simple educated guesses for what might be the proper scaling. These guesses are then tested against experimental data to judge whether the guess is successful or not. This is in fact the path taken by Wang and Castillo [3], Cal, Wang, and Castillo [4], and Wang, Castillo, and Araya [5], among others. The temperature profile similarity approach by Wang, Castillo, and coworkers appears to closely parallel the velocity profile similarity approach of Castillo and George [6]. In their temperature profile approach, Wang, Castillo, and coworkers define defect profiles similarity as

$$\frac{T(x,y/\delta_s)-T_\infty}{T_s(x)} \;=\; f(y/\delta_s) \qquad \text{for all y,} \tag{1}$$

where $f$ is some profile function of the dimensionless height, where the length scaling parameter is $\delta_s$, and where the temperature scaling parameter is $T_s$. They found that a temperature scaling parameter based on the thermal displacement thickness, the boundary layer thickness, and the temperature difference resulted in similarity behavior in a number of experimental datasets. The scaled curves plotted on top of one another just as one would expect for similar behavior.

The result in some cases was impressive in the degree with which the collapse of datasets to a single profile occurred even when datasets with different pressure gradients in the flow direction are included. However, our work herein indicates that there is a problem with this approach. In the work described below we show that if the same data is replotted as scaled temperature profiles, $T(x,y/\delta_s)/T_s(x)$, instead of scaled defect profiles, the scaled temperature profiles no longer show similarity. The experimental datasets are the same, the scaling parameters are the same, yet the results are not. The defect profile is a simple DC shift of the measured profile $T(x,y)$ using one of the endpoints from the measured profile. This type of mathematical modification cannot be changing the physics of the flow situation so any scaling parameters that work for one of the profiles must work for the other.

So how does one explain defect profile similarity but not temperature profile similarity for the same dataset and scalings? Recently Weyburne [7,8] investigated the same type of paradoxical behavior for the velocity profile similarity approach that uses defect velocity profiles instead of the measured velocity profiles. For similarity of the outer region of the velocity profiles, Castillo and George [6] base their claims of similarity on examination of plots of the defect profiles defined as $u_e(x)-u(x,y)$ where $u(x,y)$ is the velocity in the flow direction (*x*-direction) and $u_e(x)$ is the corresponding velocity at the boundary layer edge.

Scaled defect profile plots from a wide variety of datasets did indeed show collapse toward a single curve indicative of similar behavior. However, Weyburne [7,8] showed that in all of the available datasets, when the same data and scaling parameters are replotted as scaled velocity profiles, they no longer show similar behavior. This paradoxical behavior lead to the discovery of what Weyburne termed "true similarity" and "false similarity". The false similarity case occurs when defect similarity is present but the quantity $u_e(x)/u_s(x)$, where $u_s(x)$ is the scaling parameter, is not constant as required by the flow governing equation approaches for defect profiles developed by Rotta [9], Townsend [10], Castillo and George [6], and Jones, Nickels, and Marusic [11].

In what follows we show that the same type of false similarity appears to be present in the approach by Wang, Castillo, and coworkers [3-5] for temperature profile similarity. Side by side plots of experimental profiles and defect profiles indicate the same paradoxical problem; defect similarity is present but temperature profile similarity is not. A detailed explanation of the problem and the implications are reviewed in the Discussion Section.

## 2. Reduced Temperature Profile Similarity

To understand this false similarity behavior, we start by defining the traditional definition of similarity of the temperature profile. For the temperature profile on a plate depicted in Fig. 1, similarity is defined as the case when two temperature profiles taken from different stations along the flow differ only by simple scaling parameters in $y$ and $T(x,y)$. Take the length scaling parameter as $\delta_s$ and temperature scaling parameter as $T_s$. These scaling parameters can vary with the flow direction (*x*-direction) but not in the direction perpendicular to the wall (*y*-direction). The scaled temperature profile at a station $x_1$ along the wall will be similar to the scaled temperature profile at $x_2$ if

$$\frac{T(x_1,y/\delta_s)}{T_s} = \frac{T(x_2,y/\delta_s)}{T_s} \quad \text{for all y.} \tag{2}$$

Rather than deal with the experimentally measured temperature profile directly, Wang, Castillo, and coworkers [3-5] choose to parallel the velocity profile approach and work with a calculated defect profile. Thus, instead of Eq. 2, Wang, Castillo, and coworkers [3-5] definition for temperature profile similarity is given by Eq. 1. Wang and Castillo [3] considered a number of different temperature scaling parameters but their preferred parameter is what they referred to as a new outer region temperature scaling which we will call $T_{SO}$ given by

$$T_{SO}(x) = \frac{\delta_T^*(x)}{\delta_{99}(x)}(T_w - T_\infty) \tag{3}$$

where $\delta_{99}$ is the 99% boundary layer thickness and where $\delta_T^*$ is the thermal displacement thickness given by

$$\delta_T^* = \int_0^\infty dy \; \frac{T(x,y) - T_\infty}{T_w - T_\infty} \quad . \tag{4}$$

For the *y* scaling parameter they indicated that $\delta_{SO} = \delta_{99}$ is the appropriate length scale.

As part of the proof of the applicability of their new outer scaling parameter, Wang, Castillo, and Araya [5] generated plots using turbulent ZPG and APG datasets from Blackwell [12]. Using what appears to be the same exact data available in Blackwell, Kays, and Moffat [13], we reproduce Wang, Castillo, and Araya [5] Figure 3b as our Fig. 2a. (Note that our ZPG data starts at $Re_\theta$= 1204 instead of $Re_\theta$=515. We found the 515 profile did not collapse onto the other curves). The collapse of all three datasets to a single curve is impressive.

While the defect profile similarity is apparently present, we must emphasize that the actual measured data is the temperature profile. The defect profile is a calculated profile formed by subtracting off one of the endpoints of the temperature profile. There is nothing wrong with working with a calculated profile but our expectations must be that the underlying physics of the flow is not changed by generating a simple DC shift of the original dataset. Therefore the scaling parameters that work for the defect profile must necessarily work for the temperature profile. So in Fig. 2b, we replot the data from Fig. 2a as just scaled temperature profiles. While some of the profiles show similarity type behavior, the overall assessment of looking at Fig. 2b is that similarity is not present. The appearance of defect profile similarity without being accompanied by temperature profile similarity is the nature of the false similarity phenomena discussed in the Introduction.

To emphasize this paradoxical behavior, consider that there is not a mathematical reason that the defect profile in Eq. 2 must use the endpoint $T_\infty$. One should expect that the same similarity behavior if we define the defect profile similarity using any other datum from the dataset. An obvious choice is to look at the other endpoint and define similarity as

$$\frac{T_w - T\left(x_1, y/\delta_{SO}\right)}{T_{SO}\left(x_1\right)} = \frac{T_w - T\left(x_2, y/\delta_{SO}\right)}{T_{SO}\left(x_2\right)} \quad \text{for all y.} \tag{5}$$

To test this, the data and scaling parameters from Fig. 2a are plotted as the Eq. 5 defect in Fig. 2c. As in Fig. 2b, it does not appear that overall similarity is present.

In another paper, Cal, Wang, and Castillo [4] offered additional experimental proof for the scaling suggested by Wang and Castillo [3]. In Fig. 3a we reproduce Fig. 3c from Cal, Wang, and Castillo [4]. It consists of turbulent boundary layer data from Orlando, Kays, and Moffat [14] plotted using the Wang and Castillo [3] stretching parameters. In Fig. 3b and Fig. 3c we replot the same data as scaled temperature profiles and the scaled wall based defect profile. As was the case in Fig. 2, ones overall assessment is that while defect profile similarity is present, temperature profile similarity is not.

As further confirmation, we offer additional experimental proof in Fig. 4. In Fig. 4a the FPG turbulent boundary layer data from Kearney, Moffat, and Kays [15] is plotted using the Wang and Castillo [3] stretching parameters. This data consists of the non-blowing 091069 (red lines) and 070869 (blue lines) datasets. In Fig. 4b and Fig. 4c we replot the same data as scaled temperature profiles and the scaled wall based defect profile. As was the case in Figs. 2 and 3, ones overall assessment is that while defect profile similarity is present, temperature profile similarity is not.

**5. Discussion**

In looking at the boundary layer flow depicted in Fig. 1 there are two quantities of interest: the velocity profile $u(x,y)$ and the temperature profile $T(x,y)$. There has been an extensive research effort to understand the physical processes that control the evolution of the profiles as one moves down the plate in the flow direction. Similarity research, a form of dimensional analysis, is a powerful tool in that it tells us what physical properties of the fluid are important in determining the flows evolution. This non-dimensionalizing process often takes the form of using a defect profile instead of $u(x,y)$ or $T(x,y)$. This is particularly true of the search for similarity scaling parameters for the turbulent boundary layer flow. So it is not surprising that Wang, Castillo, and coworkers [3-5] presented scaled defect profile plots to determine if similarity was present or not. Working with defect profiles is perfectly acceptable as long as long as the outcome of that research applies to the experimentally measured temperature profile. The ultimate goal must be to understand the forces controlling the measured profile, not some shifted profile since to do so obfuscates what is happening in the boundary layer region.

Unfortunately, this does not appear to be the current state of the research into similarity of the turbulent boundary layer. It is now the case that the defect profile has become paramount. This is an astonishing state of affairs. How is it that the majority of the turbulent boundary layer research community has come to the point that they believe it is not the experimentally measured profiles that are important but rather some calculated profile based on the measured profile? This has been taken to the point that to even suggest the similarity study of the experimentally measured turbulent profiles is considered wrong. One would expect that the literature would fully explain the logic or theoretical basis for this current state of affairs. However, a check of the literature reveals there are no logical or theoretical justifications that have been published on this topic. In fact, Weyburne [16] has indicated that the flow governing equation approaches to similarity criterion for the defect and velocity profiles are equivalent.

To understand how the defect profile has become dominant, we need to look is back at Clauser's [17] early experimental work on scaling of the turbulent boundary layer velocity profile. Clauser is apparently the first to use the defect profile in the search for similarity in experimental turbulent datasets. His Figs. 2 and 3 compare a number of experimental datasets plotted as velocity profiles and defect profiles [17]. Whereas the defect profiles from different groups collapsed to a single universal profile, the same data plotted as velocity profiles did not. Following closely after Clauser's experimental observation, Rotta [9] and Townsend [10] developed defect profile based theoretical treatments for the study of turbulent boundary layer similarity. This association of the defect profile with the turbulent boundary layer has been reinforced by the extensive work that had occurred on the turbulent boundary layers inner region. The foundation for the Logarithmic Law of the Wall that describes the velocity profile behavior in the near wall region is the von Kármán's [18] analytical expression based on the defect profile. Given Clauser's early experimental observation and the lack of any theoretical evidence to the contrary, it has been the Clauser view that the defect profile is "discovering" similarity behavior that the velocity profile did not find that has been the accepted paradigm. Subsequent searches for similarity scaling parameters for the turbulent boundary layer have adapted the use of the defect profile as a means of discovering similar behavior. Hence, the accepted similarity definition for the velocity profile has become the defect profile based equation given by

$$\frac{u_e - u(x, y/\delta_s)}{u_s(x)} = h(y/\delta_s) \quad , \tag{6}$$

where *h* is some profile function of the dimensionless height.

As one would expect the outward appearance of the work generated using the present paradigm looks reasonable. It is not until you look more closely that one begins to see a major flaw. The flaw can be traced to Eq. 6. One of the theoretically derived requirements for defect profile similarity is that $u_e/u_s$ must be a constant. This requirement was obtained in the studies of Rotta [9] (see his Eq. 14.3) Townsend [10] (see his Eq. 7.2.3), Castillo and George [6] (see their Eq. 9), and Jones, Nickels, and Marusic [11] (see their Eq. 3.9, a2+a4). Neither they, nor anyone else, made the connection that this condition along with Eq. 6 necessarily requires that defect and velocity profile similarity must occur simultaneously. You cannot have defect profile similarity unless velocity profile similarity is first present. No one made the connection because they had visual proof from Clauser's Fig. 2 and 3 that in fact it was possible to have defect similarity but not velocity profile similarity. So this rather obvious observation about $u_e/u_s$ and similarity equivalence has gone unnoticed. It has continued to go unnoticed because of the way researchers have searched for similarity in turbulent boundary layers. Using educated guesses for the stretching parameters, researches have simply plotted the data as scaled defect profiles and visually checked the plotted curves for similarity. If the plots show similar behavior then the thinking was that there is no need to do any additional checking. So it has been the case that no one bothered to check whether the quantity $u_e/u_s$ is a constant as required by theory until Weyburne [7,8] did so. Weyburne made the observation that a simple check to insure this condition is met is to plot the data as both scaled defect profiles and scaled velocity profiles. If both sets of curves show similar behavior, then the condition is insured and similarity has been identified. In those datasets that Weyburne checked from Castillo and George [6] and others [8], this was not the case. Defect profile similarity was present but velocity profile similarity was not. Hence the requirement that $u_e/u_s$ be a constant was not met for those datasets. The revelations by Weyburne make it clear that the way the defect profile has been used to study similarity is flawed. It reinforces what should have been recognized earlier and that is that simply DC shifting the experimental data does not change the physics of the situation. Hence our focus should be to understand how the experimentally observed velocity (or temperature) profile is evolving as we move along the plate in the flow direction.

Having discussed the problem with velocity defect similarity, we now turn our attention back to temperature profile similarity. As we mentioned above, Wang, Castillo, and coworkers [3-5] approach to temperature profile similarity closely parallels the velocity profile approach of Castillo and George [6] right down to the use of the defect profile. So the question is whether the defect profile problem with the velocity profile also exists for the temperature profile. The answer from observing Figs. 2-4 is a resounding yes; the problem exists. The scaled defect profiles show similarity but the scaled temperature profiles do not. This is the same data and the same scaling parameters but with different results.

What should be important here is that the plotted scaled temperature profiles do not show similarity. However, given this strong adherence in the literature to the defect profile, it is worth trying to understand how it is that the remarkable results evident in Figs. 2a, 3a, and 4a have failed to insure temperature profile similarity. To explain how we can have defect profile similarity but not temperature profile similarity, let us start by reiterating the ultimate goal of our research effort. In spite of the prevailing paradigm believing otherwise, the ultimate goal must be to understand the evolutionary behavior of the experimentally observed profiles along the plate. It is okay to work with the defect profile but that requires we also ensure that the quantity $T_\infty / T_s(x)$ is a constant at each measurement station showing similarity. That insures that at least the endpoint of the scaled temperature profiles will be equal. Examination of the near boundary layer edge in Figs. 2b, 3b, and 4b indicate that this ratio is not constant. Looking at the APG m=-0.2 data (red line) in Fig. 2b, for example, the span between the lowest data point to the largest is about 15% compared to the mean value in the boundary layer edge region. This region is precisely the region where the scaled profile becomes $T_\infty / T_s(x)$. Hence the reason that scaled temperature profiles in Fig. 2b do not show similarity is that the quantity $T_\infty / T_s(x)$ is not a constant.

If the quantity $T_\infty / T_s(x)$ is not a constant then why does the defect profile data show similarity? To understand why this is the case, consider that the new scaling $T_{SO}(x)$ is proportional to the thermal displacement thickness $\delta_T^*$ given by Eq. 4. This parameter is called the thermal displacement thickness. However, looking at Eq. 4 it is also the area under the defect profile. Now for defect profile similarity we must have the case that the area under the scaled defect profile plotted versus the scaled height must be equal at each measurement station along the plate [19]. Mathematically, this is given by

$$a(x) = \int_0^{h/\delta_s} d\left\{\frac{y}{\delta_s}\right\} \frac{T(x, y/\delta_s) - T_\infty}{T_s(x)} \tag{7}$$

where $h$ is located far into the bulk fluid. Now if we use $T_s = T_{SO} = (T_w - T_\infty)\delta_T^* / \delta_{99}$ and $\delta_s = \delta_{99}$, then Eq. 7 becomes

$$\begin{aligned}
a(x) &= \int_0^{h/\delta_s} d\left\{\frac{y}{\delta_s}\right\} \frac{T(x, y/\delta_s) - T_\infty}{T_s(x)} \\
a(x) &= \int_0^{h/\delta_s} d\left\{\frac{y}{\delta_{99}}\right\} \frac{T(x, y/\delta_s) - T_\infty}{(T_w - T_\infty)\delta_T^* / \delta_{99}} \\
a(x) &= \frac{1}{\delta_T^* / \delta_{99}} \frac{1}{\delta_{99}} \int_0^h dy \, \frac{T(x, y/\delta_s) - T_\infty}{T_w - T_\infty} \\
a(x) &= \frac{1}{\delta_T^*} \int_0^h dy \, \frac{T(x, y/\delta_s) - T_\infty}{T_w - T_\infty} \\
a(x) &= 1 \ .
\end{aligned} \tag{8}$$

Hence the area under the scaled defect profile plotted versus the scaled height will be normalized and equal to one using $T_s = T_{SO}$ and $\delta_s = \delta_{99}$. So as long as the profile shape does not change drastically as one moves down the plate, the scaled defect profile curves will therefore plot on top of one another whether similarity is present or not. Having equal areas is a necessary but not sufficient condition to insure similarity [19]. Again we emphasize that what is important is not the appearance of the defect profile plots but whether or not temperature profile similarity is present or not.

## 6. Conclusions

The similarity scaling approach for turbulent boundary layers using the defect profile was shown to fail to insure temperature profile similarity for certain datasets. Based on the above analysis, it seems likely that this type of false similarity may be present any time the defect profile is used alone to infer temperature profile similarity. Since the ultimate goal is to investigate similarity of the temperature profile of turbulent boundary layers, it is suggested to forgo the use of defect profile and concentrate on the temperature profile itself.

## Acknowledgement

The author would like to acknowledge the support of the Air Force Research Laboratory and Gernot Pomrenke at AFOSR. In addition, the author would like to thank the experimentalists for making their datasets available for inclusion in this manuscript.

## REFERENCES

[1] L. Prandtl, "*Über Flüssigkeitsbewegung* bei sehr kleiner Reibung," *Verhandlungen des Dritten Internationalen Mathematiker-Kongresses in Heidelberg 1904*, A. Krazer, ed., Teubner, Leipzig, pp. 484–491, 1905.

[2] O. Reynolds, "An experimental investigation of the circumstances which determine whether the motion of water in parallel channels shall be direct or sinuous, and of the law of resistance in parallel channels," Philosophical Transactions of the Royal Society of London. **174**, 935-982(1883).

[3] X. Wang and L. Castillo, "Asymptotic Solutions in Forced Convection Turbulent Boundary Layers," J. Turbul., **4**, 1(2003).

[4] R. Cal, X. Wang, and L. Castillo, "Transpired Turbulent Boundary Layers Subject to Forced Convection and External Pressure Gradients," Journal of Heat Transfer, **127,** 194(2005).

[5] X. Wang, L. Castillo, and G. Araya, "Temperature Scalings and Profiles in Forced Convection Turbulent Boundary Layers." Journal of Heat Transfer, **130**, 021701 (2008).

[6] L. Castillo and W. George, "Similarity Analysis for Turbulent Boundary Layer with Pressure Gradient: Outer Flow," AIAA J. **39**, 41 (2001).

[7] D. Weyburne, "The Prevalence of Similarity of the Turbulent Wall-bounded Velocity Profile," arXiv:1412.5129 [physics.flu-dyn], 2015.

[8] D. Weyburne, "A Cautionary Note on the Zagarola and Smits Similarity Parameter for the Turbulent Boundary Layer," arXiv:1507.06951 [physics.flu-dyn], 2015.

[9] J. Rotta, "Turbulent Boundary Layers in Incompressible Flow," Prog. Aeronaut. Sci. **2**, 1(1962).

[10] A. Townsend, *The Structure of Turbulent Shear Flows*, 2nd ed. (Cambridge Univ. Press, Cambridge, U.K., 1976).

[11] M. Jones, T. Nickels, and I. Marusic, “On the asymptotic similarity of the zero-pressure-gradient turbulent boundary layer,” J. Fluid Mech. **616,** 195(2008).

[12] B. Blackwell, “The Turbulent Boundary Layer on a Porous Plate: An Experimental Study of the Heat Transfer Behavior with Adverse Pressure Gradients,” Ph.D. thesis, Stanford University, Palo Alto, 1972.

[13] B. Blackwell, W. Kays, and R. Moffat, "The Turbulent Boundary Layer on a Porous Plate: An Experimental Study of the Heat Transfer Behavior With Adverse Pressure Gradients," Report No. HMT-16, Thermosciences Division, Dept. of Mechanical Engineering, Stanford Univ., Stanford, CA, 1972, available as NASA NTRS report NASA-CR-130291.

[14] A. Orlando, W. Kays, and R. Moffat, ‘‘Turbulent Transport of Heat and Momentum in a Boundary Layer Subject to Deceleration, Suction, and Variable Wall Temperature,’’ Report No. HMT-17, Dept. of Mech. Eng., Stanford University, Stanford, CA, 1974, available as NASA NTRS report NASA-CR-139655.

[15] D. Kearney, R. Moffat and W. Kays, “The Turbulent Boundary Layer: Experimental Heat Transfer with Strong Favorable Pressure Gradients and Blowing,” Report No. HMT-12, Thermosciences Division, Dept. of Mechanical Engineering, Stanford Univ., Stanford, CA, 1970, available as NASA NTRS report NASA-CR-110653.

[16] D. Weyburne, “Are Defect Profile Similarity Criteria Different Than Velocity Profile Similarity Criteria for the Turbulent Boundary Layer?” ArXiv:1510.05588 [physics.flu-dyn], 2015.

[17] F. Clauser, “The turbulent boundary layer in adverse pressure gradients,” J. Aeronaut. Sci. **21**, 91(1954).

[18] Thv. Kármán, “Mechanische Ähnlichkeit und Turbulenz,” Nachr Ges Wiss Göttingen, Math Phys Klasse, 58(1930) {see also NACA-TM-611, 1931}.

[19] D. Weyburne, “Similarity of the Temperature Profile formed by Fluid Flow along a Wall,” ArXiv:1603.05062 [physics.flu-dyn], 2016.

**Figures**

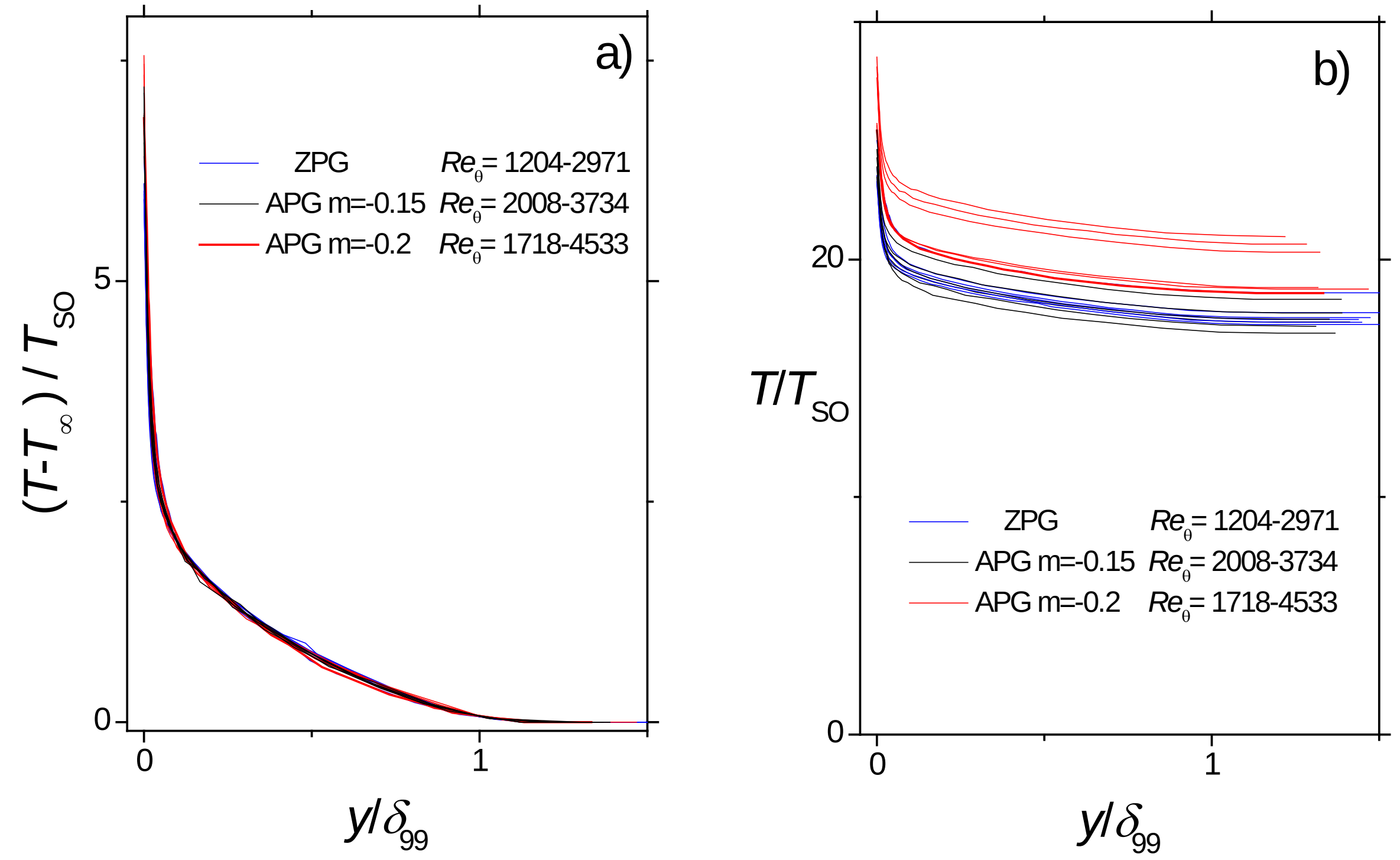

Fig. 2: Blackwell, Kays, and Moffat [13] data plotted as a) defect profiles, b) temperature profiles, and c) alternate defect profiles. All are plotted using $T_{SO} = (T_w - T_\infty)\delta_T^* / \delta_{99}$ .

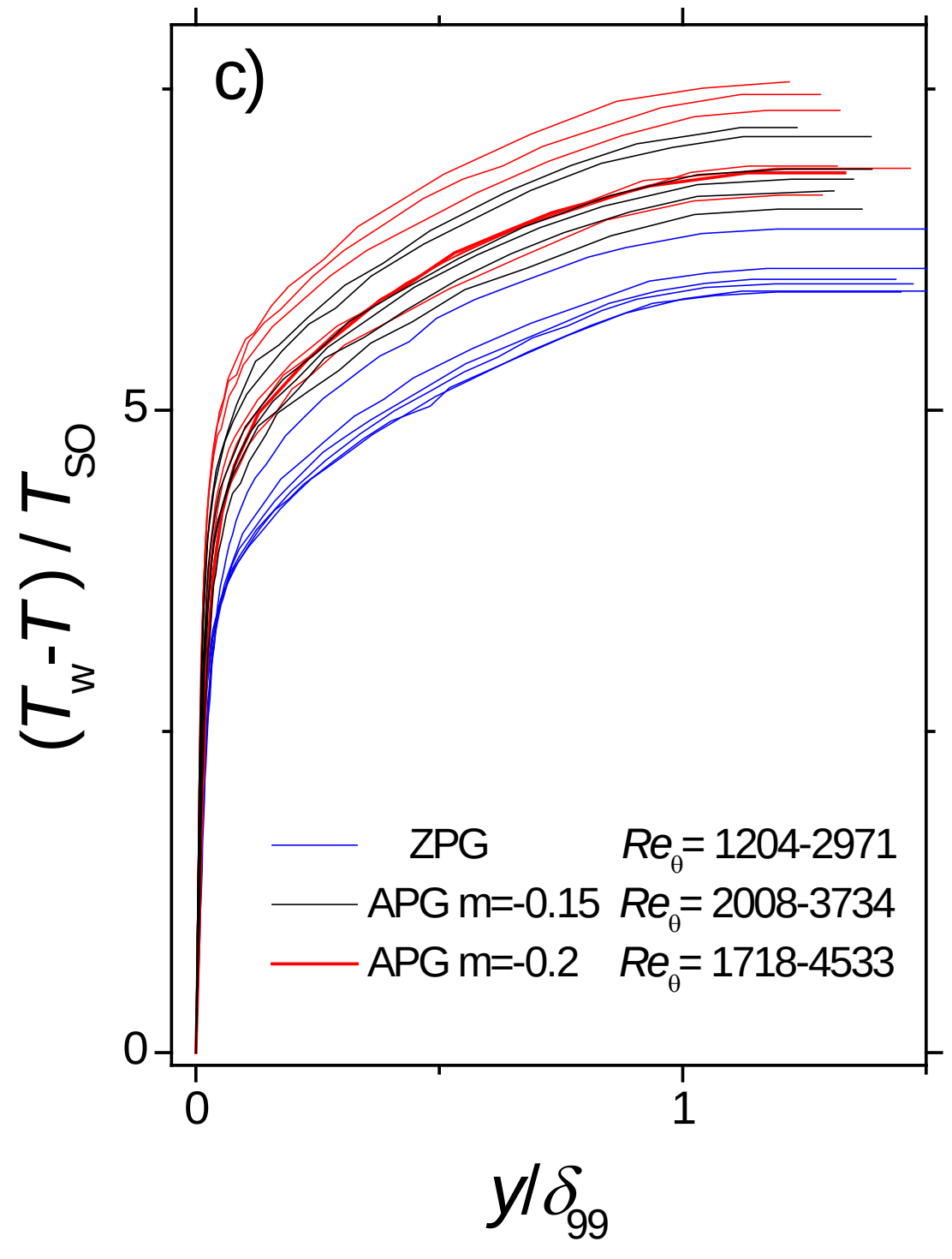

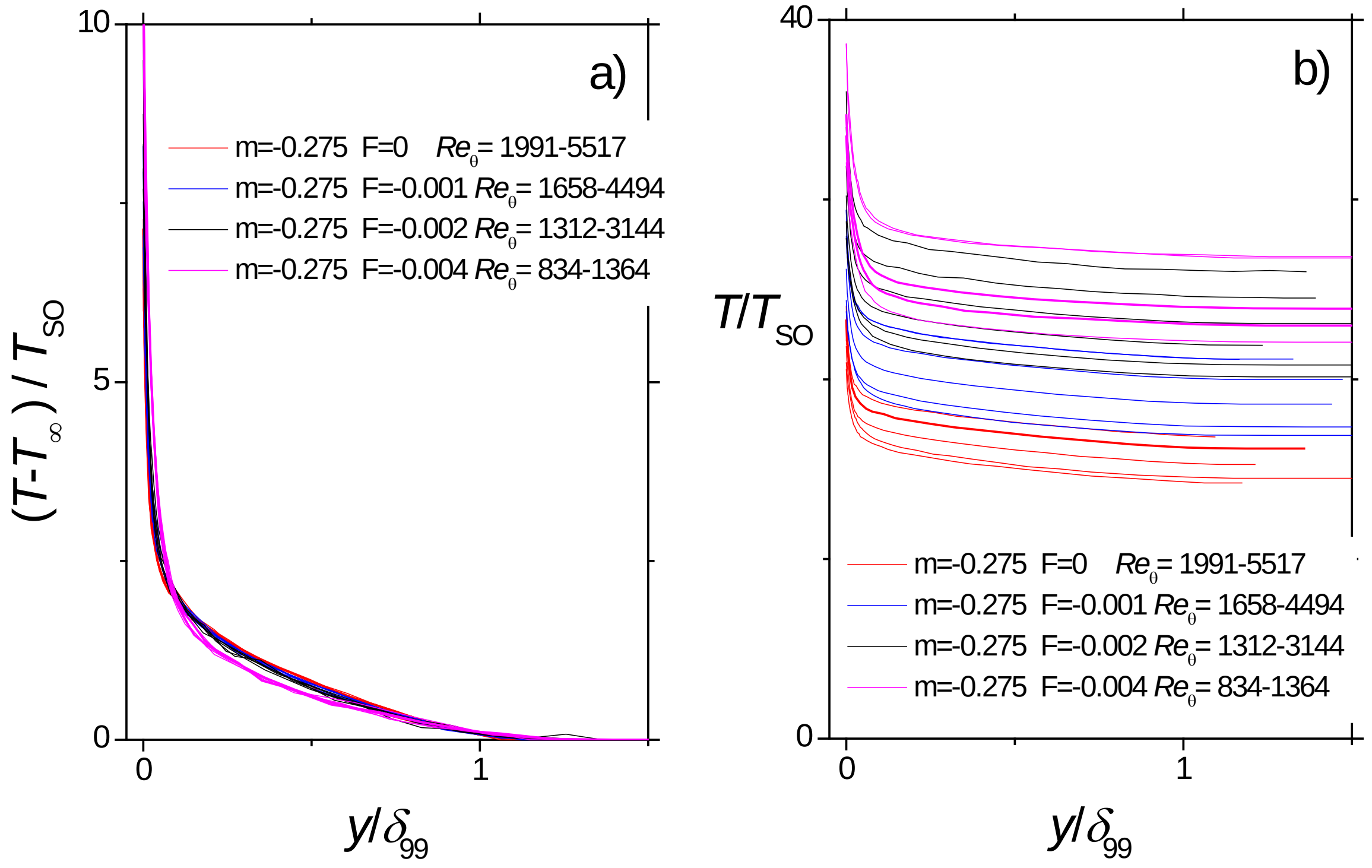

Fig. 3: Orlando, Kays, and Moffat [14] APG turbulent data plotted as a) defect profiles, b) temperature profiles, and c) alternate defect profiles. All are plotted using $T_{SO} = \left(T_w - T_\infty\right)\delta_T^* / \delta_{99}$ .

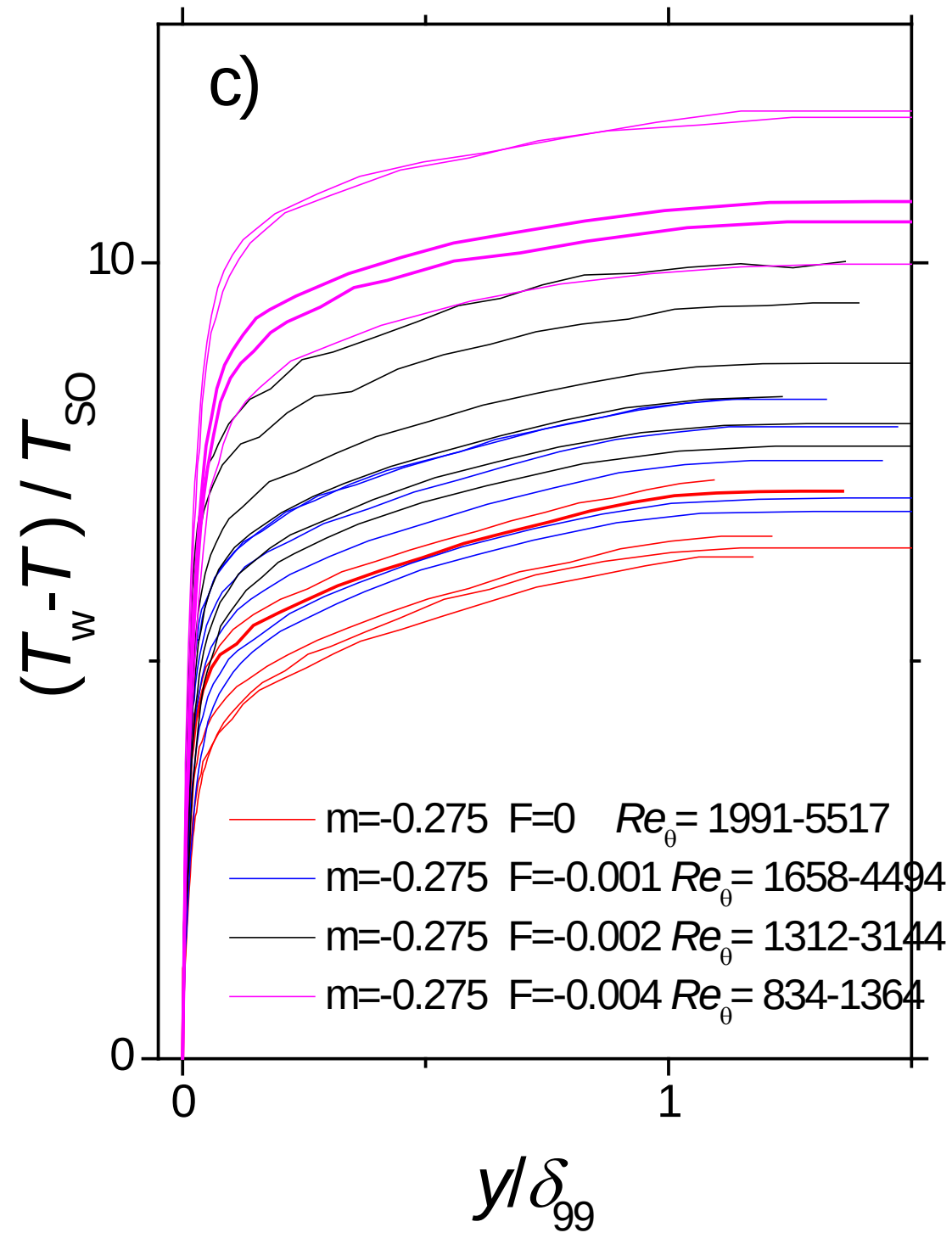

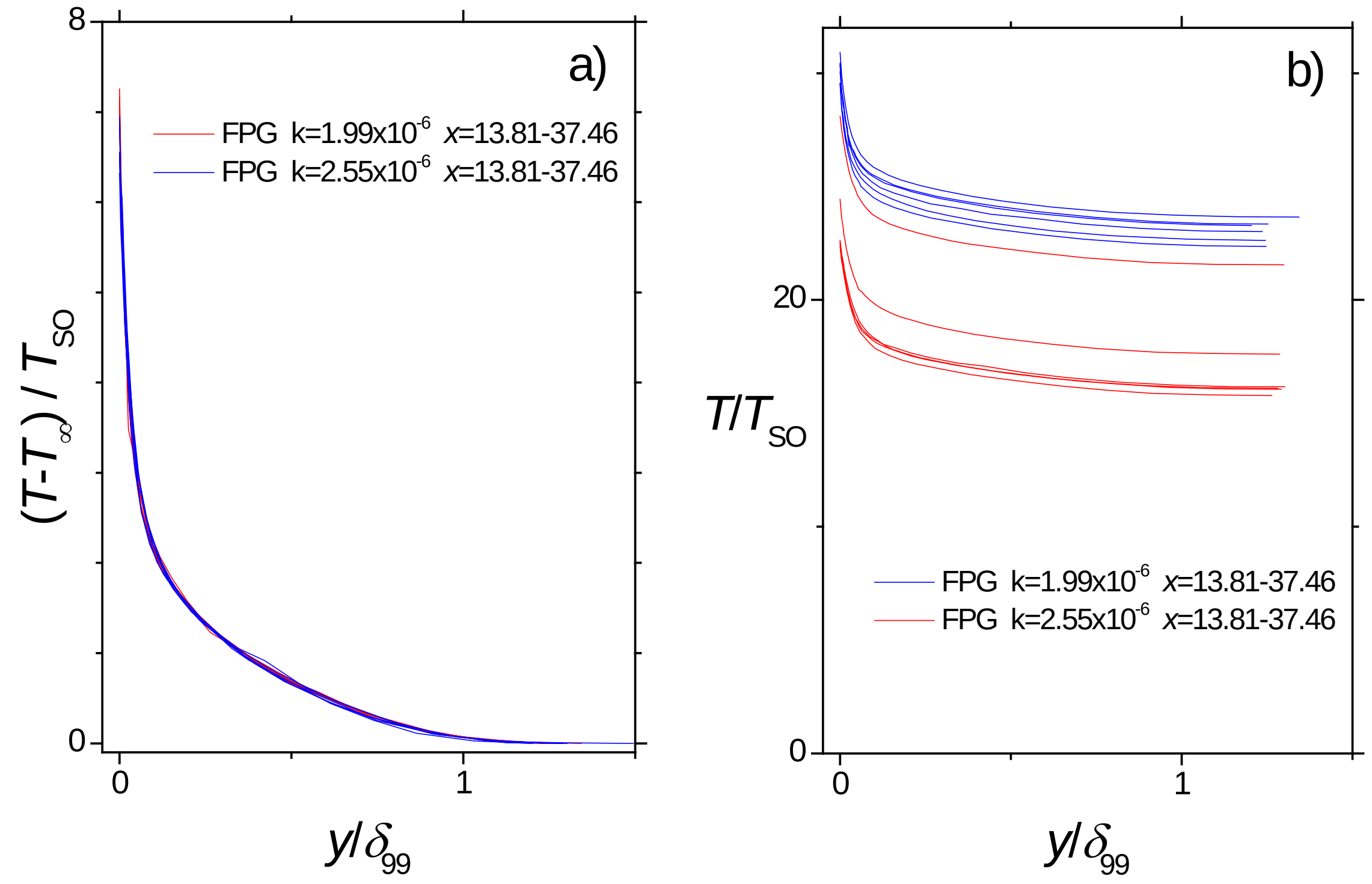

Fig. 4: Kearney, Moffat, and Kays [15] FPG turbulent data plotted as a) defect profiles, b) temperature profiles, and c) alternate defect profiles. All are plotted using $T_{SO} = \left(T_w - T_\infty\right)\delta_T^* / \delta_{99}$ .

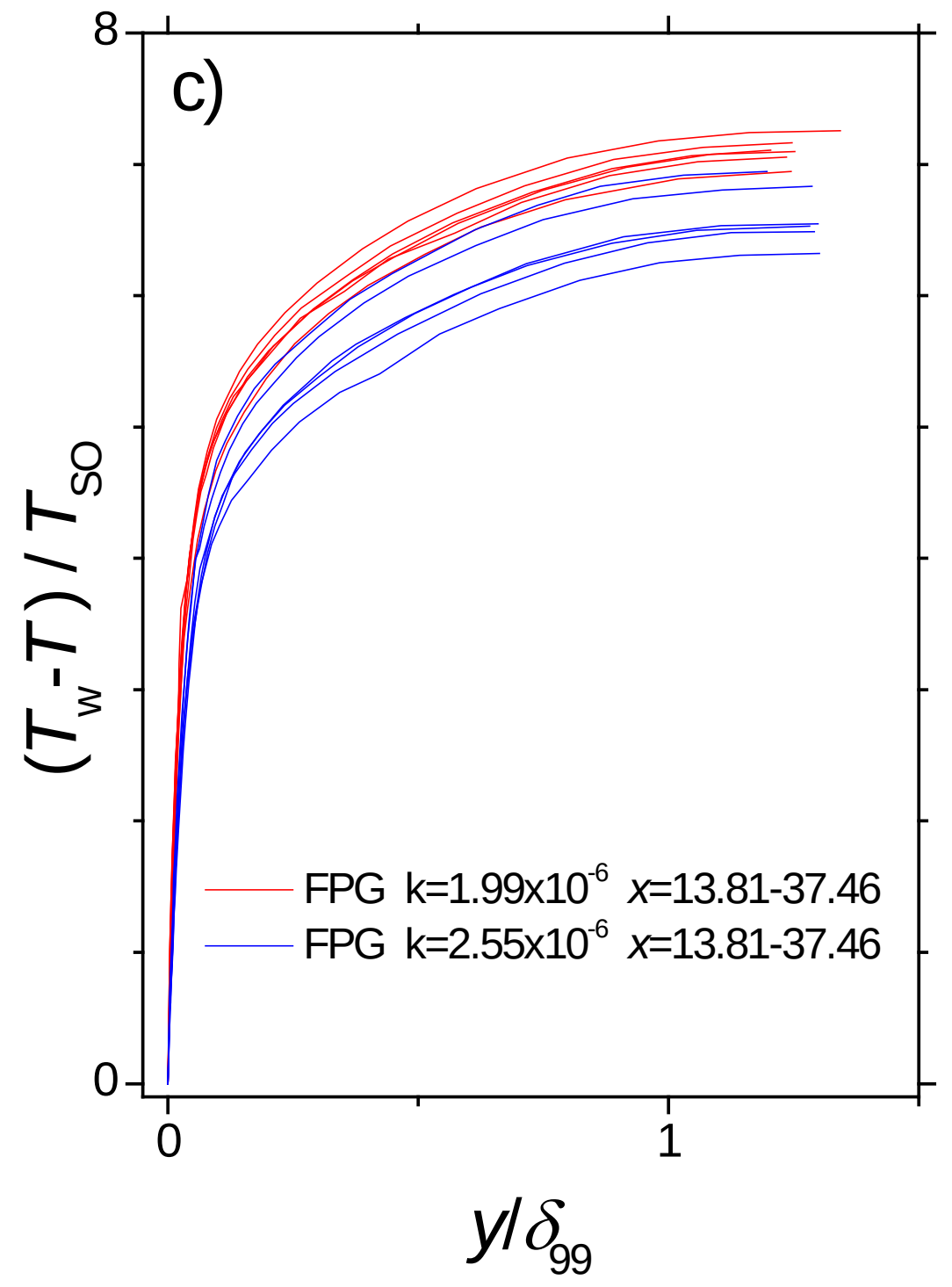